\documentclass[aps,prl,twocolumn,groupedaddress,showpacs,floatfix]{revtex4-1}
\usepackage[T1]{fontenc} 
\usepackage{lmodern}
\usepackage{graphicx}
\usepackage{color}
\usepackage{amsmath}
\usepackage[overload]{empheq}

\begin{document}

\author{Matthias Dauth$^{1,2}$, Fabio Caruso$^3$, Stephan K\"ummel$^1$, and Patrick Rinke$^2$}
\affiliation{$^1$ Theoretical Physics IV, University of Bayreuth, D-95440 Bayreuth, Germany}
\affiliation{$^2$ COMP Department of Applied Physics, Aalto University, P.O. Box 11100, Aalto FI-00076, Finland}
\affiliation{$^3$ Department of Materials, University of Oxford, Parks Road, Oxford OX1 3PH, United Kingdom}

\title{Piecewise linearity in the $GW$ approximation for accurate quasiparticle energy predictions}

\begin{abstract}
We identify the {\it deviation from the straight line error} (DSLE) -- i.e., the spurious non-linearity of the total energy as a function of fractional particle number
-- as the main source for the discrepancy between experimental vertical ionization energies and theoretical quasiparticle energies, as obtained from the $GW$ and $GW$+SOSEX approximations to many-body perturbation theory (MBPT). 
For self-consistent calculations, we show that $GW$ suffers from a small DSLE.
Conversely, for perturbative $G_0W_0$ and $G_0W_0$+SOSEX calculations the DSLE depends on the starting point. We exploit this starting-point dependence to reduce (or completely eliminate) the DSLE. We find that the agreement with experiment increases as the DSLE reduces. DSLE-minimized schemes, thus, emerge as promising avenues for future developments in MBPT. 
\end{abstract}


\maketitle

Electronic structure theory has developed into an essential tool in material science, because it offers a parameter free, quantum mechanical description of solids, molecules, and nano-structures. This success is due to the continuous development of electronic structure methods such as density-functional theory (DFT)~\cite{KS65} and many-body perturbation theory (MBPT) in the $GW$ approximation~\cite{Hedin1965}. This development is guided in part by comparison with experimental reference data and in part by exact constraints, imposed by the theoretical framework itself. We will demonstrate in this work that invaluable insight into the $GW$ approach can be gained from such an exact constraint. The $GW$ method has long been heralded as the method of choice for band gaps and band structures of solids and quasiparticle spectra of molecules and nano-structures~\cite{Aulbur,rubioRMP,RQNFS05}. Yet, its accuracy is not always satisfying and the starting-point dependence in the perturbative $G_0W_0$ variant can be very pronounced~\cite{RQNFS05,Fuchs2007,Marom2012a}. Here we will show that the accuracy of $GW$ is closely related to the exact constraint of piecewise linearity of the total energy~\cite{PPLB82}. Its violation gives rise to the \emph{deviation from the straight line error} (DSLE),  which has been extensively studied in DFT~\cite{YangScience,manyelsiperdew,manyelsiyang,ZY98,kraislerprl,Vlcek2015}. The starting-point dependence in $G_0W_0$ can then be exploited to minimize the DSLE, which uniquely defines the optimal starting point.

In 1982 Perdew {\it et al.} showed that the total energy of a quantum mechanical system has to change linearly with respect to the  fractional removal (or addition) of an electron~\cite{PPLB82}
\begin{equation}
 E(f) = (1-f) E(N_0-1) + f E(N_0).
 \label{eq:straight_line}
\end{equation}
Here, $N_0$ is the number of electrons in the neutral system and $E(N_0)$ the associated total energy.  $E(N_0-1)$ is the total energy of the singly ionized system and $f$ varies in the interval $[0,1]$. This piecewise linearity condition was initially derived in the context of DFT, but applies to any total energy method. 

\begin{figure}[b]
\includegraphics[width=0.5\textwidth]{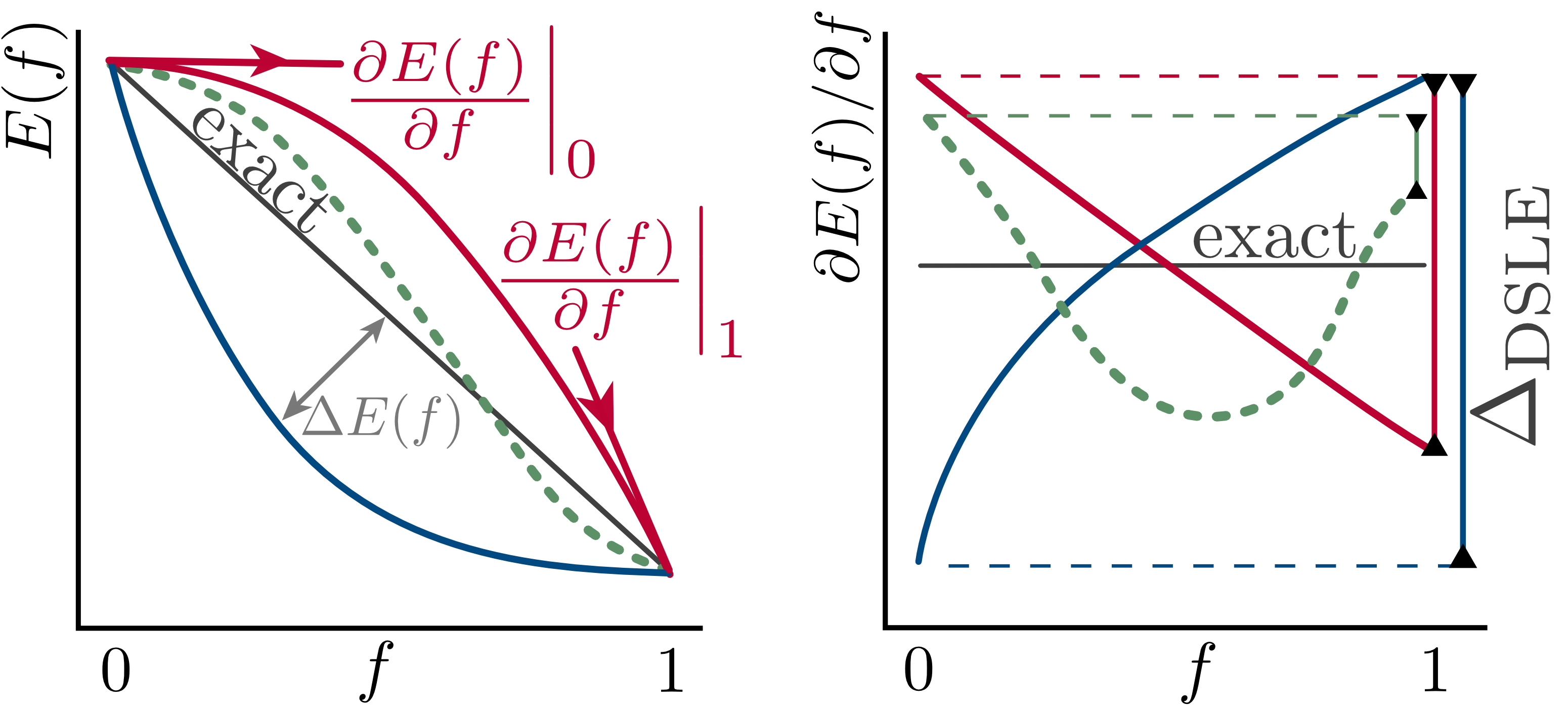}
 \caption{
Schematic representation of the DSLE for total energies $E$ (left) 
and their derivatives $\partial E / \partial f$ (right) as a function of the occupation number $f$.}
\label{fig:dsle-schem}
\end{figure}
In the following, we introduce a formal definition of the deviation from linearity at fractional occupation numbers
that employs only quantities directly accessible by quasiparticle energy calculations, that is, the 
ionization potential (IP) and the electron affinity (EA). 
The linearity condition in Eq.~(\ref{eq:straight_line}) implies that the first
derivative of the total energy with respect to the fractional occupation number $f$ (i.e., $\partial E / \partial f$) 
should be constant and that it exhibits discontinuities at integer occupations ($f=0$ and $f=1$).
Additionally, $\partial E / \partial f$ equals the electron removal energy $E(N_0) - E(N_0-1)$ or, equivalently, 
the energy for adding an electron to the positively charged system. 
Therefore, the vertical IP of the neutral system with $N_0$ electrons can be expressed as: 
\begin{equation}\label{eq:ip}
\left. \frac{\partial E(f)}{\partial f}\right|_{1} = E(N_0) - E(N_0-1) = -\rm{IP}(N_0). 
\end{equation}
A similar relation holds for the electron affinity of the cation (EA$_{\rm c}$): 
\begin{equation}\label{eq:ea}
\left. \frac{\partial E(f)}{\partial f}\right|_{0} = E(N_0) - E(N_0-1) = -\rm{EA}_{\rm c}(N_0-1).
\end{equation}
Equations~(\ref{eq:ip}) and (\ref{eq:ea}) illustrate that ${\rm IP} = {\rm EA}_{\rm c}$  in an exact theory. We thus define the difference between IP and EA$_{\rm c}$ as the DSLE
\begin{equation}\label{eq:delta}
\Delta_{\mathrm{DSLE}} = {\rm EA}_{\rm c}(N_0-1) - {\rm IP}(N_0). 
\end{equation}
$\Delta_{\mathrm{DSLE}} = 0$ is a necessary condition for piecewise linearity.
However, in an approximate treatment of electronic exchange and correlation, as e.g. in $GW$, IP and EA$_{\rm c}$ may differ. A non vanishing $\Delta_{\mathrm{DSLE}}$ indicates
a curvature in the total energy versus fractional electron number curve, as illustrated in Fig.~\ref{fig:dsle-schem}, causing an erroneous deviation from the straight line~\cite{Atalla2014,Bruneval2009}. 

\begin{figure}
\includegraphics[width=0.5\textwidth]{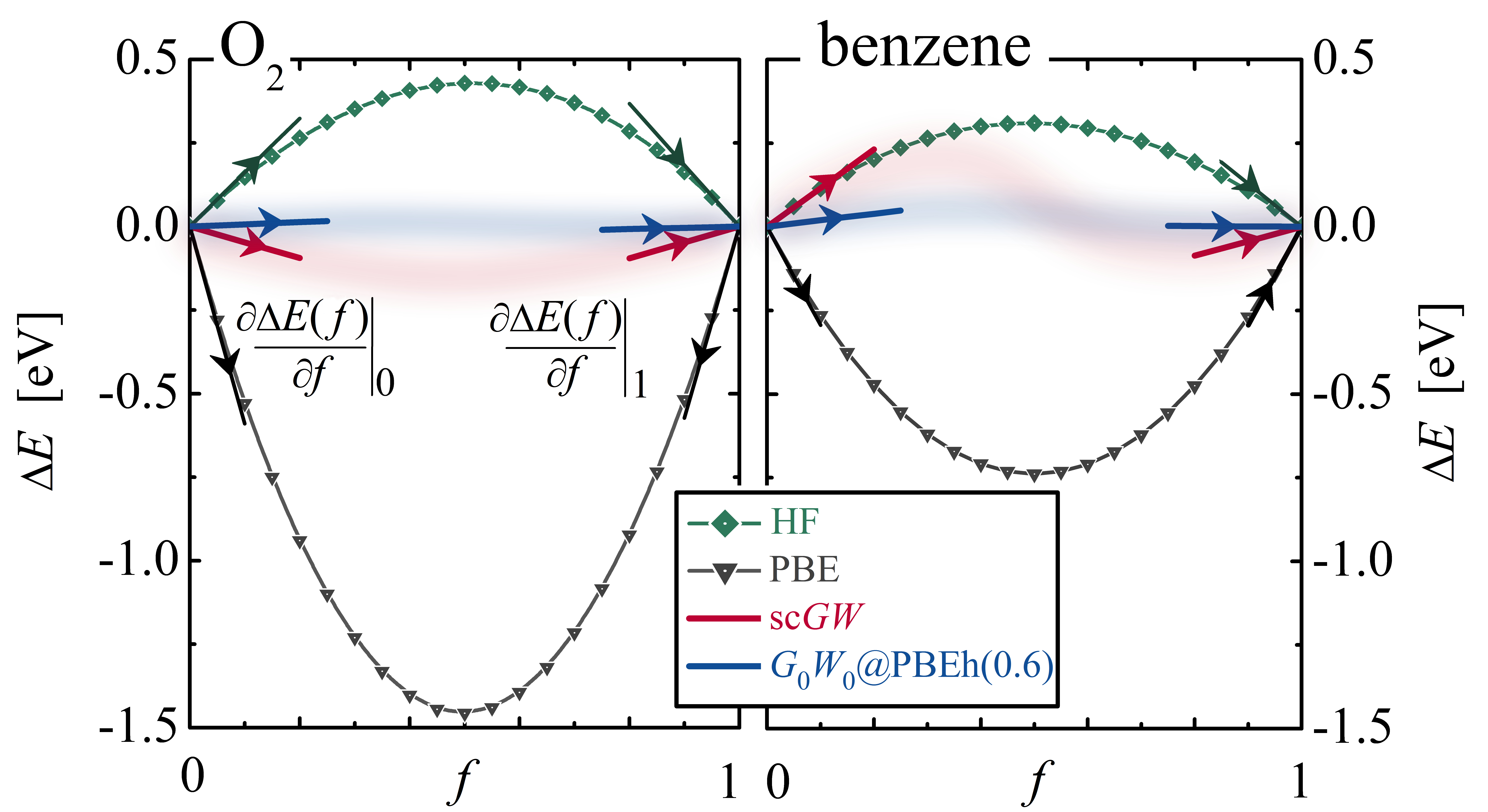}
 \caption{Deviation from the straight $\Delta E(f)$ for O$_2$ (left) and benzene (right). The slopes of $\Delta E$ are indicated by arrows which we obtained by evaluating Eq.~(\ref{eq:DdeltaE}) with either sc$GW$ quasiparticle and total energies, or RPA@PBEh(0.6) total and $G_0W_0$@PBEh(0.6) quasiparticle energies. The expected behavior of $\Delta E$ for sc$GW$ and $G_0W_0$@PBEh(0.6) are depicted as blurred curves.}
\label{fig:Denergy}
\end{figure}

The DSLE is most easily seen in the deviation from the straight line 
\begin{equation}
 \label{eq:D_etot}
 \Delta E(f) = E(f) -E_{\mathrm{lin}}(f),
\end{equation}
where, following Eq.~(\ref{eq:straight_line}), $E_{\mathrm{lin}}(f)$ is the straight line between $E(N_0)$ and $E(N_0-1)$. We will first examine the DSLE and Eq.~(\ref{eq:D_etot}) for different DFT functionals before proceeding to our $GW$ analysis.
In DFT, common (semi-)local functionals typically exhibit a convex curvature and suffer from a large DSLE, whereas Hartree-Fock (HF) is concave with a moderate DSLE~\cite{YangScience,manyelsiperdew,manyelsiyang,ZY98,kraislerprl,Vlcek2015}. We show this tendency in terms of $\Delta E(f)$ for the examples of the O$_2$ and the benzene molecule in Fig.~\ref{fig:Denergy}.
\footnote{
We have performed all calculations with the Fritz-Haber-Institut {\it ab initio} molecular simulations (FHI-aims) package~\cite{Blum2009,Ren2012}.
We used Tier 3 numeric atom-centered orbital basis functions~\cite{Ren2012} for self-consitent $GW$~\cite{Caruso2012a,Caruso2013}, otherwise the Tier 4 basis sets augmented by Gaussian aug-cc-pV5Z basis functions~\cite{Ren2012,GaussBasisSet}}
To quantify the DSLE by means of Eq.~(\ref{eq:delta}) we use the eigenvalue of the highest molecular orbital (HOMO) of the neutral system ($\epsilon^{\rm H}_{N_0}$) for IP$(N_0)$ and the eigenvalue of the lowest unoccupied molecular orbital (LUMO) of the cation ($\epsilon^{\rm L}_{N_0-1}$) for EA$_{\rm c}(N_0-1)$~\cite{AB85,PPLB82,Janak1978,WYang2012}.
For O$_2$, PBE gives  $\Delta_{\mathrm{DSLE}} = \epsilon^{\rm H }_{N_0} - \epsilon^{\rm L}_{N_0-1} = 11.6~\mathrm{eV}$, which agrees with the pronounced convexity observed in Fig.~\ref{fig:Denergy}. For benzene, the DSLE in PBE reduces to half the size ($\Delta_{\mathrm{DSLE}} = 5.9~\mathrm{eV}$), which is also apparent from the maximal extent of $\Delta E(f)$ in Fig.~\ref{fig:Denergy}. 
Conversely, HF exhibits  a concave DSLE manifested in  $\Delta_{\mathrm{DSLE}}=-3.5~\mathrm{eV}$ for O$_2$ and $\Delta_{\mathrm{DSLE}} = -2.5~\mathrm{eV}$ for benzene.
All in all, the magnitude of $\Delta_{\mathrm{DSLE}}$ can be taken as a measure for the severity of the DSLE, whereas the sign indicates the curvature. 
A positive value of $\Delta_{\rm DSLE}$ corresponds to a concave and a negative sign to a convex curvature. Convexity gives rise to a delocalization of the electron density and concavity to an overlocalization~\cite{YangScience}.

We now move on to discuss the DSLE in the $GW$ approximation. In $GW$, the total energy is only available at integer occupation numbers, because an ensemble generalization to  fractional particle numbers is not straightforward. However, we can still estimate the basic shape of the total energy curve by invoking only observables at integer particle numbers. If, in addition to the quasiparticle energies, we also have access to the total energy at integer particle numbers, we can calculate the slopes of $\Delta E$
\begin{equation}\label{eq:DdeltaE}
 \partial \Delta E \slash \partial f = \partial E \slash \partial f - \partial E_{\rm lin} \slash \partial f.
\end{equation}
For the reference straight line we use $\partial E_{\rm lin}\slash \partial f =  E(N_0-1) -E(N_0)$, and $\partial E \slash \partial f$ gives us the $GW$ quasiparticle energies. Equation (\ref{eq:DdeltaE}) then becomes
\begin{subequations}
\label{eq:Dderiv}
  \begin{align}[left ={\frac{\partial \Delta E}{\partial f } = \empheqlbrace}]
     \epsilon_{N_0}^{\rm H} +  E(N_0-1) -E(N_0) \; \;  & \text{for} \; N_0 \label{eq:DderivH} \\
     \epsilon_{N_0-1}^{\rm L} +  E(N_0-1) -E(N_0) \; \; & \text{for} \; N_0-1, \label{eq:DderivL}
  \end{align}
\end{subequations}
where we calculate the total energies $E(N_0-1)$ and  $E(N_0)$ with fully self-consistent $GW$ (sc$GW$)~\cite{Caruso2012a,Caruso2013}. The resulting slopes are shown as arrows in Fig.~\ref{fig:Denergy}.

Among the different $GW$ flavors only sc$GW$, in which the Dyson equation is solved 
iteratively, gives results that are independent of the starting point~\cite{Caruso2012a,Caruso2013}. Therefore, sc$GW$ provides an unbiased assessment of the DSLE. Most importantly, in sc$GW$ also the ground state is treated at the $GW$ level and the sc$GW$ density does not inherit  the (de-)localization error of the starting point, as is the case in $G_0W_0$ calculations.

Figure~\ref{fig:Denergy} illustrates that also sc$GW$ violates the straight line condition due to the approximate nature of the self-energy.
Evaluating Eq.~(\ref{eq:delta}) with the sc$GW$ quasiparticle HOMO and LUMO energies, we obtain $\Delta_{\mathrm{DSLE}} = 0.9~\mathrm{eV}$ for O$_2$. This indicates convexity, which is, however, much less pronounced than in PBE. The slopes at the two end-points of the sc$GW$ curve confirm the convex behavior, because they are pointing in different directions, i.e., they have different signs. For our second example, benzene, the behavior is markedly different. The signs of $\partial \Delta E \slash \partial f$ are equal at both ends of the occupation interval. To connect both endpoints -- schematically sketched by the blurred curve in Fig.~\ref{fig:Denergy} -- $\Delta E$ inevitably has to cross zero at some point. Hence, we expect the $\Delta E$ curve to be divided into two regimes. Beginning from the cation, the positive slope at $N_0-1$ gives rise to a concave DSLE. When we approach the $N_0$ electron limit, the positive slope at $N_0$ requires a convex curvature. For benzene, the absolute value of the slope at $N_0-1$ is higher than at the other end of the interval. As a result, we expect the concave deviation on the $N_0-1$ side to be more pronounced than the convex part. This is also reflected by a negative $\Delta_{\mathrm{DSLE}} = -0.7~\mathrm{eV}$, which we associate with a concave DSLE.

Our benzene example illustrates that we can in principle encounter systems that are not DSLE-free, although $\Delta_{\rm DSLE}=0$. In these cases $\partial \Delta E \slash \partial f$ provides additional information on the total energy curve. If $\partial \Delta E \slash \partial f=0$ at $N_0-1$ and $N_0$, we can ensure that the total energy follows exactly the straight line in the vicinity of $N_0-1$ and $N$. Hence, we expect the method to  be DSLE-free. These two conditions are generally applicable to any electronic-structure method that provides access to total and quasiparticle energies. 

\begin{figure}
\includegraphics[width=0.5\textwidth]{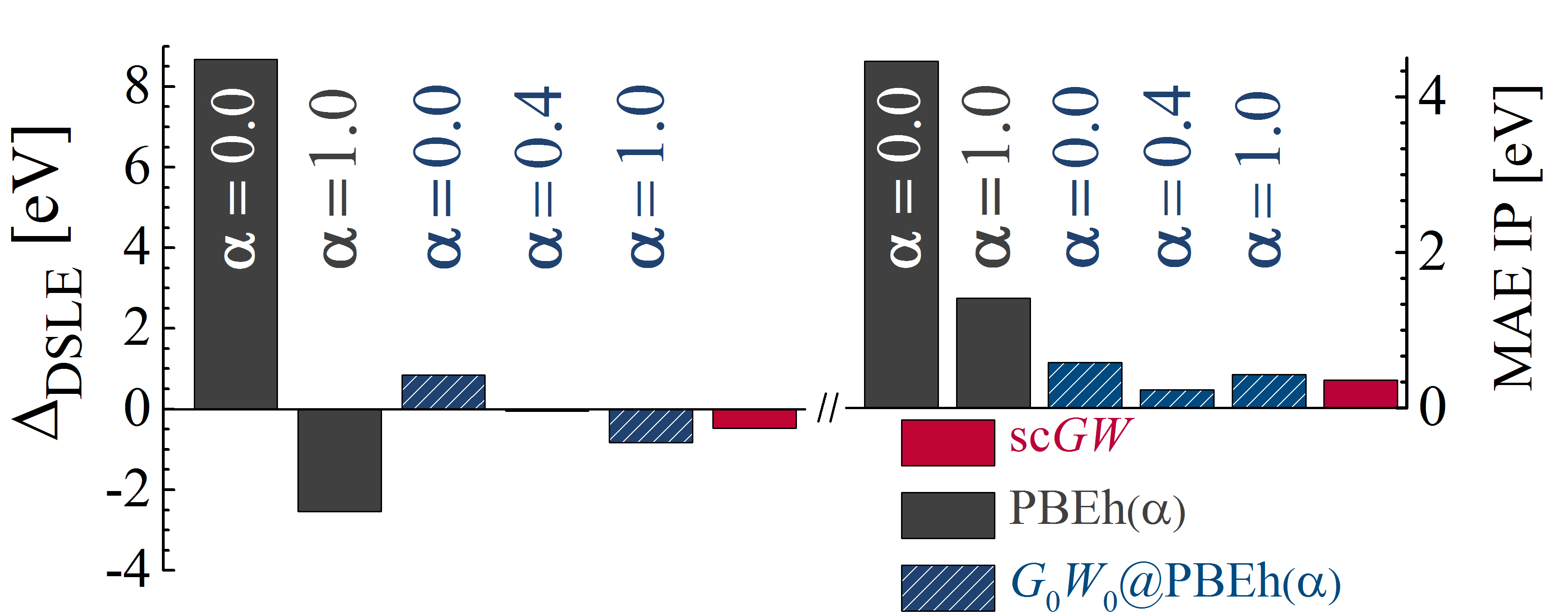}
 \caption{
Left panel: Average $\Delta_{\rm DSLE}$ for the  ${\rm G2}_{\rm ip}$ set.
Right panel: Mean average error (MAE) of the ionization potential. 
}
\label{fig:dsle-histo}
\end{figure}

To provide a comprehensive assessment of the DSLE in the $GW$ approximation, 
we have further performed calculations for a 
benchmark set consisting of 48 atoms and molecules selected from the 
quantum chemical G2 ion test set~\cite{Curtiss1997,Ren2012} 
based on the availability of experimental vertical IPs~\footnote{We excluded 
the free H and He-atom from benchmark set presented in Ref.~\cite{Ren2012} as the $G_0W_0$ 
quasiparticle LUMO energies of the cations reside in a completely empty spin channel.} -- 
referred to as ${\rm G2}_{\rm ip}$ in the following~\cite{suppl}. 
For sc$GW$ we find a mean DSLE of $\bar{\Delta}_{\mathrm{DSLE}} = -0.5~\mathrm{eV}$. The average over the absolute 
$\bar{\Delta}_{\mathrm{DSLE}}$ amounts to 0.9~eV. The sc$GW$ DSLE is thus much smaller than that of PBE and HF (Figs.~\ref{fig:Denergy} and~\ref{fig:dsle-histo}). Our results provide quantitative evidence that sc$GW$ predominantly exhibits a 
concave DSLE. More generally, the slopes at $N_0$ show a small deviation from zero, $\partial \Delta E \slash \partial f = 0.14~\mathrm{eV}$. Conversely, the slope at  $N_0-1$ is higher with a mean value of $\partial \Delta E \slash \partial f = 0.61~\mathrm{eV}$.
Out of the 48 systems of the ${\rm G2}_{\rm ip}$ set, 28 of them exhibit simultaneously concave and convex curvatures, as in the benzene case. Overall, our results suggest that fully self-consistent $GW$ has the tendency to over-localize electron density, which is consistent with previous work~\cite{Bruneval2009} on the quasiparticle self-consistent $GW$ approach~\cite{Schilfgaarde2006}. 

Most commonly, $GW$ is not carried out fully self-consistently, but applied in first-order perturbation theory ($G_0W_0$). This introduces  a dependence on the reference ground state encoded in $G_0$. Logically, also the DSLE should depend on the chosen starting point.

For O$_2$, $G_0W_0$ calculations based on orbitals and eigenvalues from PBE ($G_0W_0$@PBE) yield 
an IP of $-11.6$~eV, which differs from the A$_{\rm c}$ of the cation ($-13.0$~eV). 
Thus, $G_0W_0$@PBE also violates the straight line condition, as quantified through
Eq.~(\ref{eq:delta}), which yields $\Delta_{\mathrm{DSLE}} = 1.4~\mathrm{eV}$.
The positive value of $\Delta_{\mathrm{DSLE}}$ indicates a convex total energy at fractional particle numbers as in PBE, albeit an order of magnitude smaller than in PBE. 
Conversely, we find $\Delta_{\rm DSLE} = -0.6$~eV with opposite sign if we use $G_0W_0$ based on the PBE hybrid 
functional PBEh($\alpha$)~\cite{PBE0,PBE96} with $\alpha= 1$. 

 \begin{figure}
\includegraphics[width=0.5\textwidth]{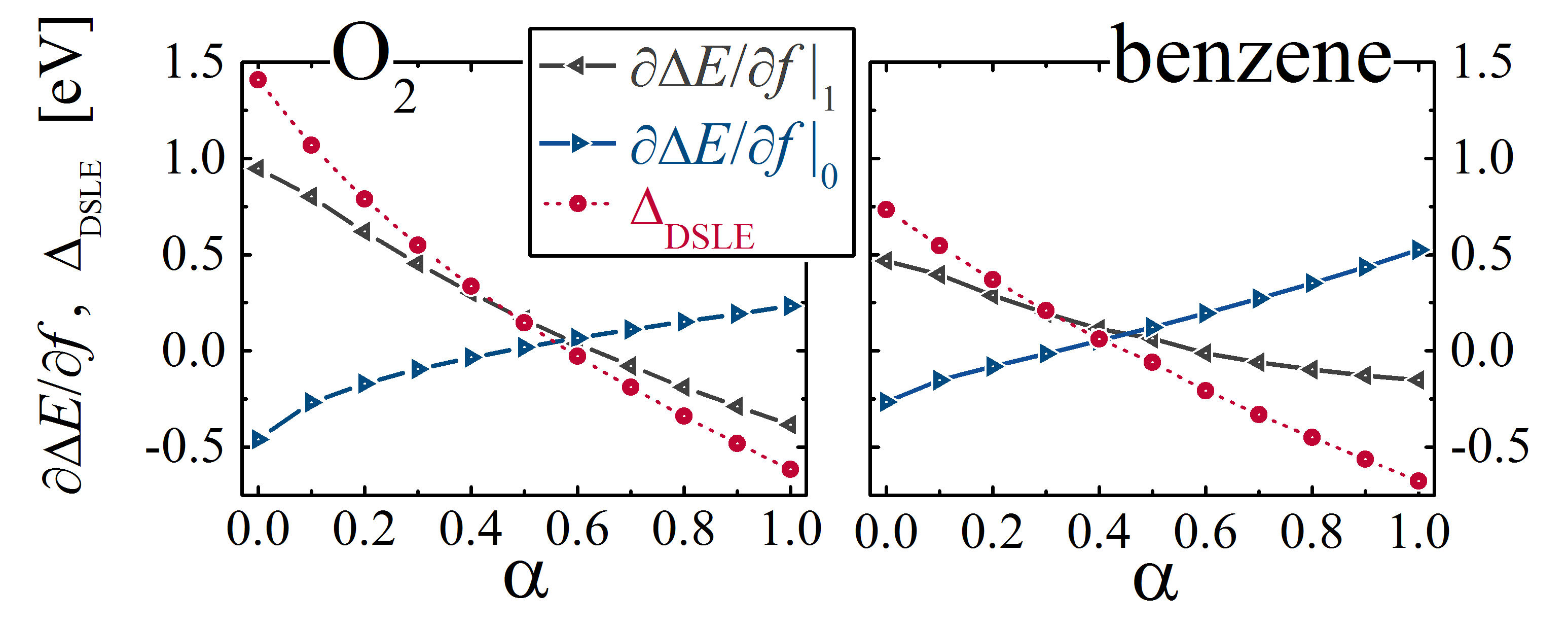}
 \caption{$\Delta_{\rm DLSE}$ (red), $\partial \Delta E \slash \partial f$  for O$_2$ and benzene at $N_0-1$ (blue) and $N_0$ (black) particles for $G_0 W_0$@PBEh($\alpha$) and RPA@PBEh($\alpha$) total energies.}
 \label{fig:g0w0_o2_benz}
 \end{figure}

Now one could ask if our DSLE definition may prove useful for the design of novel DSLE-free approaches for quasiparticle energy calculations. Since in our two examples $\Delta_{\rm DSLE}$ changes sign for $\alpha=0$ and $\alpha=1$, it is conceivable to postulate that an intermediate, optimal $\alpha$ exists for which the DSLE in $G_0W_0$ is eliminated or at least considerably reduced. To test this postulate, we evaluate the DSLE for several $\alpha$ values in the range $[0,1]$. As illustrated in the lower panel of Fig.~\ref{fig:g0w0_o2_benz}, an increasing $\alpha$ gradually decreases $\Delta_{\rm DSLE}$  for O$_2$ and benzene. At an optimal $\alpha$ of $\approx$~0.4 for benzene and $\approx$~0.6 for O$_2$ the DSLE vanishes. Beyond this point,  $\Delta_{\rm DSLE}$ becomes increasingly negative with a further increase in $\alpha$, indicating an increasingly concave total energy curve.

In analogy to DFT and sc$GW$, we can support the $G_0W_0$ $\Delta_{\rm DSLE}$ results by examining the slopes of $\Delta E$ as defined in Eq.~(\ref{eq:DdeltaE}). In $G_0W_0$ the quasiparticle energies are equivalent to the derivative of the total energy in the random phase approximation (RPA) with respect to the particle number~\cite{Wang2010}. We show $\partial \Delta E \slash \partial f$ for $N_0-1$ and $N_0$ as a function of $\alpha$ for O$_2$ and benzene in the upper panel of Fig.~\ref{fig:g0w0_o2_benz}.
Both molecules show the trend expected from the $\Delta_{\rm DSLE}$ calculations. Beginning at small $\alpha$, the slopes support  convexity because $\Delta E$ falls away from $N_0-1$ and rises again approaching $N_0$. Conversely, for large $\alpha$, the signs of the slopes are reversed, implying concavity. Both slopes approach zero around $\alpha=0.6$ for O$_2$ and $\alpha=0.4$ for benzene, which is consistent with the $\alpha$ values for which $\Delta_{\rm DSLE}$ vanishes. Figure~\ref{fig:Denergy} shows the expected behavior of the total energy for $\alpha=0.6$.

\begin{figure}
\includegraphics[width=0.5\textwidth]{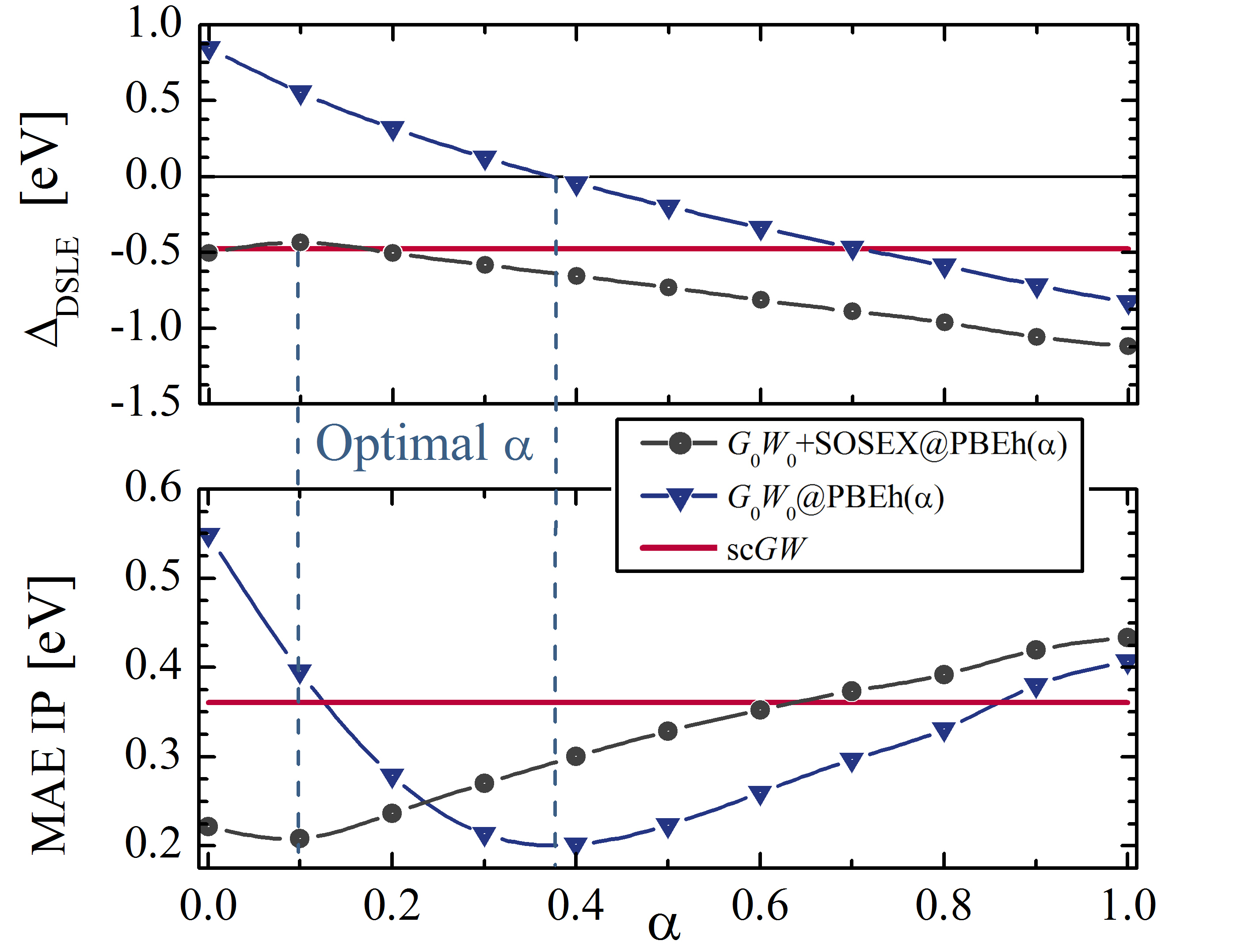}
 \caption{Upper panel:
 Average $\Delta_{\rm DSLE}$ for the G2$_{\rm ip}$ set computed with sc$GW$, $G_0W_0$, and $G_0W_0+$SOSEX,
 as a function of the exact exchange parameter $\alpha$ employed in the PBEh($\alpha$) starting point.
 Lower panel:
 Mean absolute error of the predicted ionization potential depending on the starting point. }
 \label{fig:dsle+ip}
 \end{figure}

Next, we return to our  benchmark set of 48 molecules. As illustrated in the upper panel of Fig.~\ref{fig:dsle+ip}, $G_0W_0$ exhibits the largest DSLE, when starting from PBE. The DSLE decreases for increasing $\alpha$ and approaches a negative DSLE in the limit of 100\% HF exchange that is almost as large in absolute value as for $G_0 W_0$@PBE. The optimal $\alpha$ that eliminates the DSLE again amounts to $\approx$~0.4. We also determined the optimal $\alpha$ values for each molecule, individually. Here, too, the optimal $\alpha$ amounts to $\alpha \approx 0.4$ on average. 

We now establish a correlation between the DSLE and the agreement with experimental reference data for ionization potentials. For all $M$ molecules of the ${\rm G2}_{\rm ip}$ test set, we evaluated the $G_0W_0$@PBEh($\alpha$) HOMO energies for different $\alpha$ and calculated the mean absolute error (${\rm MAE}\equiv \sum_{i=1}^{M} |\epsilon_i^{\rm H} - {\rm IP}_i^{\rm exp}| / M$)
with respect to the experimental vertical IPs~\cite{nist}.
As shown in the lower panel of Fig.~\ref{fig:dsle+ip}, the MAE exhibits a minimum at $\alpha \approx 0.4$ when $G_0W_0$@PBEh($\alpha$) becomes DSLE-minimized. At this point, the  MAE amounts to 0.20~eV \footnote{At the optimal starting point for each system, we observe the same MAE for the ${\rm G2}_{\rm ip}$ set.}, which can be seen as the intrinsic accuracy of $G_0W_0$ based on global hybrid functionals. With increasing DSLE we observe a concomitant increase of the MAE.

As the minimum in the $G_0W_0$ MAE curve is rather shallow, hybrid functional starting points with 30\%-50\% HF exchange produce reasonable MAEs. This is consistent with the  empirical findings of Bruneval~\cite{Bruneval2012} and Marom {\it et al.}~\cite{Marom2012a} 
and the work of K\"orzd\"orfer {\it et al.}~\cite{Korzdorfer2012c}.

Consistent with the analysis above, the MAE of sc$GW$ (0.36~eV) is slightly higher than for DSLE-minimized $G_0W_0$. However, the MAE of the $\Delta_{\rm scf}$IPs, i.e., the IP from the total energy difference, in sc$GW$ amounts to only 0.22~eV, and is thus comparable to DSLE-minimized $G_0W_0$. Our results strongly suggest that the DSLE is largely responsible for the discrepancies between $GW$ quasiparticle energies and experimental ionization energies.

Finally, we illustrate that the concept of DSLE-minimized quasiparticle calculations is generally applicable and can be transferred to other self-energy approximations. Motivated by the good results of renormalized second order perturbation theory (rPT2) for electron correlation energies~\cite{Ren2013}, Ren and coworkers recently proposed a beyond-$GW$ scheme~\cite{Ren2015} that combines $GW$ with a second order screened exchange self-energy ($G_0W_0$+SOSEX). In the following, we apply our DSLE analysis to ($G_0W_0$+SOSEX)@PBEh($\alpha$) calculations for the G2$_{\rm ip}$ test set. In the upper panel of Fig.~\ref{fig:dsle+ip} we display the corresponding DSLE and in the lower panel the IP MAE as a function of $\alpha$. Compared to $G_0W_0$, the starting-point dependence is weaker and the DSLE is always negative. Also the DSLE and the MAE are minimized at smaller $\alpha$ values. This confirms our previous, heuristic findings, that rPT2 and $G_0W_0$+SOSEX perform best for starting points that are close to PBE. The smallest average deviation from the experimental IPs amounts to 0.21~eV, which is comparable to $G_0W_0$.

In conclusion, we have shown that the DSLE is a prominent source of discrepancy between experimental and theoretical vertical IPs. Through a formal definition of the DSLE for quasiparticle calculations, we show that the prominent $GW$ approach has an intrinsic DSLE of -0.5~eV and a tendency towards concavity, i.e. localization of electrons. We then establish a correlation between the DSLE and the deviation from experimental ionization energies. This allowed us to propose a recipe for obtaining DSLE-minimized approximations to many-body perturbation theory. The DSLE-minimized $G_0W_0$ and $G_0W_0$+SOSEX schemes give the best agreement 
with experimental data, as illustrated for the 48 molecules of the G2$_{\rm ip}$ test set.

\begin{acknowledgments}
 We thank Xinguo Ren for supporting us with the SOSEX calculations and Christoph Friedrich for fruitful discussions. This work was supported by the Academy of Finland through its Centres of Excellence Programme under project numbers 251748 and 284621.
S.K.\ and M.D.\ acknowledge support by Deutsche Forschungsgemeinschaft 
Graduiertenkolleg 1640 and the Bavarian State Ministry of Science, Research, and the Arts for the Collaborative Research Network Soltech
\end{acknowledgments}

\end{document}